\documentclass[a4paper,11pt]{article}
\usepackage{pos}
\usepackage{subfig}
\usepackage{diagbox}

\newcommand{\cD}{\ensuremath{\mathcal D} }
\newcommand{\cDbar}{\ensuremath{\overline{\mathcal D}} }
\newcommand{\cF}{\ensuremath{\mathcal F} }
\newcommand{\cFbar}{\ensuremath{\overline{\mathcal F}} }
\newcommand{\Ibb}{\ensuremath{\mathbb I} }
\newcommand{\bn}{\ensuremath{\mathbf n} }
\newcommand{\bmu}{\ensuremath{\mathbf \mu} }

\newcommand{\cP}{\ensuremath{\mathcal P} }

\newcommand{\cU}{\ensuremath{\mathcal U} }
\newcommand{\cUbar}{\ensuremath{\overline{\mathcal U}} }

\newcommand{\la}{\ensuremath{\lambda} }
\newcommand{\lalat}{\ensuremath{\la_{\text{lat}}} }

\newcommand{\Tr}[1]{\ensuremath{\mbox{Tr}\left[ #1 \right]} }

\newcommand{\eq}[1]{Eq.~\ref{#1}}
\newcommand{\fig}[1]{Fig.~\ref{#1}}

\newcommand{\refcite}[1]{Ref.~\cite{#1}}

\title{Lattice Studies of 3D Maximally Supersymmetric Yang--Mills}

\author*{Angel Sherletov}
\author{David Schaich}

\affiliation{Department of Mathematical Sciences, University of Liverpool, \\ Liverpool L69 7ZL, United Kingdom}

\emailAdd{A.Sherletov@liverpool.ac.uk}
\emailAdd{david.schaich@liverpool.ac.uk}

\FullConference{39th International Symposium on Lattice Field Theory \\ 8--13 August 2022 \\ Rheinische Friedrich-Wilhelms-Universit{\"a}t Bonn, Germany}

\abstract{
We present ongoing investigations of maximally supersymmetric Yang--Mills ($Q = 16$ SYM) theory
in three space-time dimensions. At low temperatures and large $N$ this theory is related to black branes in higher-dimensional
quantum gravity. Building on previous work that focused on the homogeneous `D2' phase of the theory,
we are now exploring phase transitions between this D2 phase and the localized `D0' phase.
}

\begin{document}
\maketitle

\section{Introduction and Motivation}
Supersymmetry is quite prominent in modern theoretical physics and in recent years there has been considerable interest and progress in the use of lattice field theory to study supersymmetric quantum field theories non-pertubatively --- see \refcite{Schaich:2022xgy} for a recent review.

Studying supersymmetric theories on a lattice is challenging because the discetization of space-time breaks the super-Poincar\'e algebra explicitly.
An approach that has been fruitful in overcoming this challenge has been to reformulate supersymmetric Yang--Mills (SYM) theories in terms of `twisted' variables. Under this formulation, a closed supersymmetry sub-algebra can be preserved at non-zero lattice spacing, hence enabling the recovery of the correct continuum limit with little to no fine-tuning.
This approach  has its limits --- in $d$ dimensions we must have a SYM theory with $Q \geq 2^d$ supersymmetries as described in \refcite{Catterall:2009it}.

In this proceedings we expand on the work done in \refcite{Catterall:2020nmn}. We present ongoing lattice studies of phase transitions in 3d SYM with $Q = 16$.
Conjectured holographic duality relates this theory to IIA supergravity~\cite{Itzhaki:1998dd}, which is compelling to test with the non-perturabtive approach of lattice field theory.
This three-dimensional case is particularly promising to study, thanks to the more modest computational costs compared to four-dimensional $\mathcal N = 4$ SYM.

We begin in the next section by briefly summarizing the lattice theory. Then, in Section~\ref{sec:expect} we discuss expectations for the theory, including those from holography. Finally, in Section~\ref{sec:results} we present preliminary results that are consistent with a phase transition, and briefly conclude in Section~\ref{sec:conc}.

\section{Three-dimensional maximal SYM on the lattice}
We obtain three-dimensional maximal SYM from the four-dimensional theory using naive dimensional reduction.
That is, we carry out four-dimensional lattice calculations with the number of sites in the $z$ direction set to $N_z=1$. The lattice action of the underlying four-dimensional theory theory is~\cite{Catterall:2020nmn}
\begin{multline}
  \label{eq:S}
  S = \frac{N}{4\lalat} \sum_{\bn} \mathrm{Tr} \Big[-\cFbar_{ab}(\bn)\cF_{ab}(\bn) + \frac{1}{2}\left(\cDbar^{(-)}_a \cU_a(\bn)\right)^2 - \chi_{ab}(\bn)\cD^{(+)}_{[a}\psi_{b]}(\bn) \\
  - \eta(\bn)\cD^{(-)}_a\psi_a(\bn) - \frac{1}{4}\epsilon_{abcde}\ \chi_{de}(\bn+\hat{\bmu}_a+\hat{\bmu}_b+\hat{\bmu}_c)\cD^{(-)}_{[c}\chi_{ab]}(\bn)\Big].
\end{multline}
Here $\{\eta, \psi_a, \chi_{ab}\}$ are $\{1, 5, 10\}$-component fermions, while the gauge field and six scalars are combined in the five-component complexified gauge links $\{\cU_a, \cUbar_a\}$ that appear in the field strength $\cF_{ab}$ and finite-difference operators $\cD_a$.
These complexified gauge links lead to U($N$) rather than SU($N$) gauge invariance~\cite{Catterall:2015ira}.

In addition we include in the lattice action two soft-supersymmetry-breaking deformations that help stabilize numerical computations.
The first is
\begin{equation}
  \label{eq:pot}
  \frac{N}{4\lalat} \mu^2 \sum_{\bn, a} \mathrm{Tr}\left[\bigg(\cU_a(\bn) \cUbar_a(\bn) - \Ibb_N\bigg)^2\right].
\end{equation}
This deformation lifts the SU($N$) flat directions.
The second is
\begin{equation}
  \label{eq:dimred}
  \frac{N}{4\lalat} \kappa^2 \sum_{\bn} \mathrm{Tr}\left[\bigg(\cU_z(\bn) - \Ibb_N\bigg)^\dag \bigg(\cU_z(\bn) - \Ibb_N\bigg)\right].
\end{equation}
This forces the trace of $\cU_z(\bn)$ to be close to $N$, explicitly breaking the center symmetry and ensuring that we implement Kaluza--Klein dimensional reduction as opposed to Eguchi--Kawai volume reduction.

The dimensionally reduced 3d theory lives on the body-centered cubic ($\mathrm{A}^*_3$) lattice with coordination number $8$, i.e., every lattice site has $8$ nearest neighbours.
We identify one of the directions as corresponding to Euclidean time by imposing thermal boundary conditions on it (periodic for the bosons and anti-periodic for the fermions).
The size of the skewed 3-torus~\cite{Catterall:2020nmn} in this direction is $r_{\beta}$, and in this work we fix the sizes of the remaining spatial dimensions to be the same, $r_x = r_y = r_L$.

These dimensionless lengths are defined as $r_L=L\lambda$ and $r_{\beta}=\beta\lambda$ where $\lambda$ is the 't~Hooft coupling and $L$ and $\beta$ are the dimensionful lengths in the spatial and temporal directions, respectively. The dimensionless quantities remain well-defined on an $N_L\times N_L\times N_t$ lattice, with $r_L=N_L\lalat$ and $r_{\beta}=N_t\lalat$, where \lalat is the dimensionless lattice 't~Hooft coupling.  The temporal extent can be thought of as an inverse dimensionless temperature $1/r_{\beta} \equiv t$.
The aspect ratio of the lattice, $\alpha \equiv r_L / r_{\beta} = L / \beta$, is set by these spatial and temporal extents.

The continuum limit is obtained by taking $N_L, N_t \to \infty$ with $\lalat\rightarrow 0$ so as to keep $\alpha$, $r_L$ and $r_B$ fixed.
To recover the correct continuum theory in this limit, we also need to send $\mu\rightarrow 0$ and $\kappa\rightarrow 0$, which we ensure by setting $\mu=\kappa=\zeta\lalat$.
In the lattice calculations presented below, we will carry out scans of the inverse dimensionless temperature $r_{\beta}$ for several values of $\zeta$ and several $N_L^2 \times N_t$ lattice volumes.

The code used for this system is the publicly available package presented in \refcite{Schaich:2014pda}.\footnote{{\tt\href{https://github.com/daschaich/susy}{github.com/daschaich/susy}}}
Several improvements and extensions to this software have recently been implemented, including an improved 4d action~\cite{Catterall:2015ira}, 0+1-dimensional matrix models~\cite{Schaich:2022duk, Dhindsa:2022uqn}, the deformation in \eq{eq:dimred}, and ongoing development of code for 3d SYM with $Q = 8$ supercharges~\cite{Sherletov:2022rnl}.
These developments will soon be presented in an update of \refcite{Schaich:2014pda}.

\section{\label{sec:expect}Theory Expectations}
\begin{figure}
  \centering
  \includegraphics[width=0.45\textwidth]{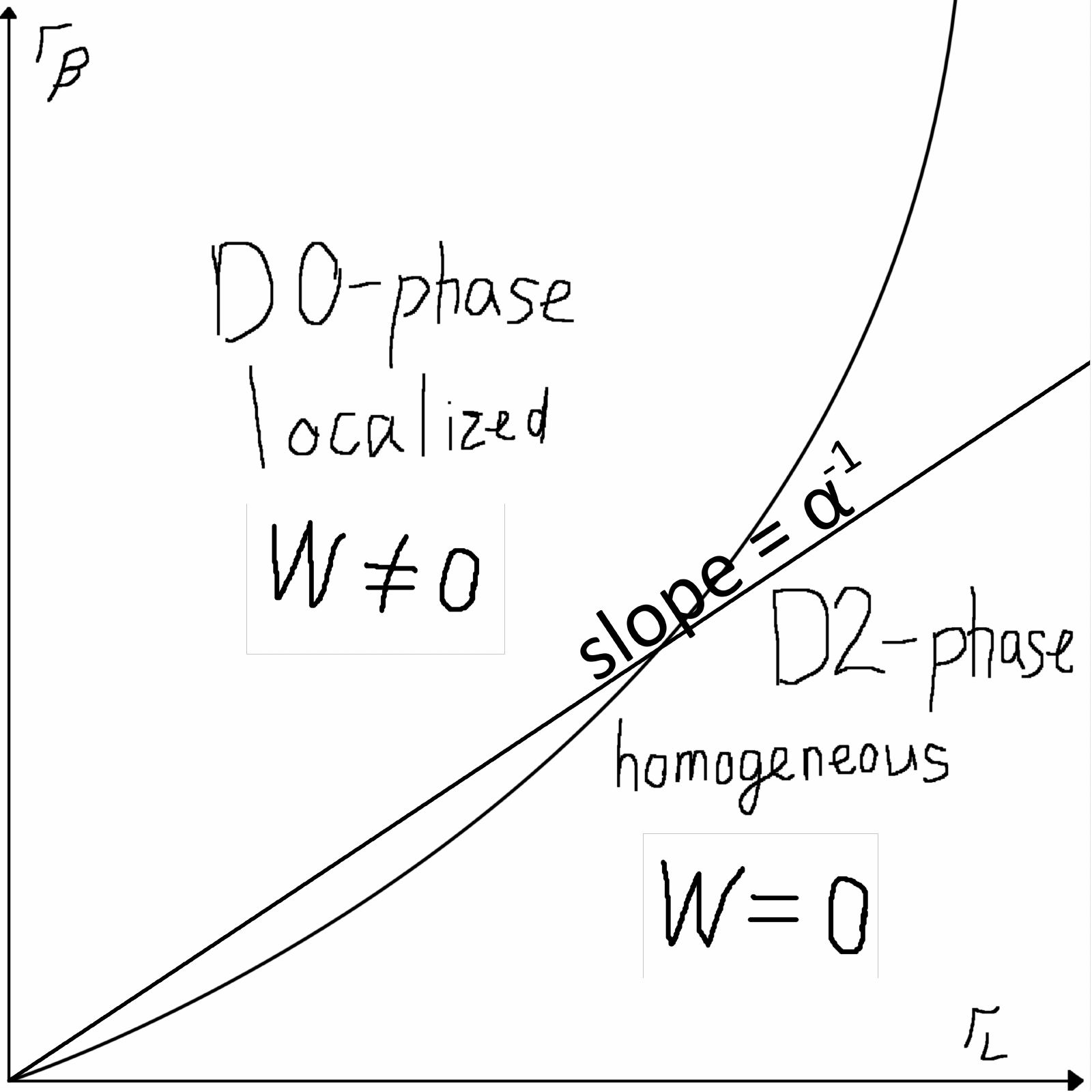}
  \caption{\label{fig:expec}A sketch of the expected phase diagram, which features spatial confinement for large $r_L$ and spatial deconfinement for small $r_L$.  The straight line corresponds to a fixed aspect ratio $\alpha = r_L / r_{\beta}$.  The curve is the holographic expectation of \eq{eq:holo}, which assumes low temperature (large $r_{\beta}$).}
\end{figure}

As in the two-dimensional case considered by \refcite{Catterall:2017lub}, there are two regimes where we have reliable theoretical expectations for the phase diagram of the 3d maximally supersymmetric theory.
First, for very small $r_{\beta}$, the very large temperature strongly breaks supersymmetry, causing the fermions to pick up large thermal masses and decouple, leaving a simpler two-dimensional gauge--scalar system.
In the opposite regime of low temperatures (large $r_{\beta}$), large-$N$ holographic duality predicts that the field theory should be well described by dual supergravity solutions.
In particular, for large $r_L$ the theory should be dual to a system of homogeneous D2 black branes, with a first-order phase transition separating this `D2 phase' from a small-$r_L$ `D0 phase' of localized black holes.
These expectations are illustrated in \fig{fig:expec} and discussed by \refcite{Morita:2014md} for a general number of dimensions.

In the field theory context, the D0 and D2 phases are distinguished by the spatial Wilson line --- the holonomy $W = \Tr{\cP e^{i \oint_L A}}$ around either spatial dimension of the 3-torus: $W \ne 0$ in the D0 phase, while the D2 phase is characterized by $W = 0$ in the large-$N$ limit.
We will refer to these phases of the field theory as `spatially deconfined' and `spatially confined', respectively, in analogy with the thermal confinement transition that would be signalled by the Polyakov loop (the holonomy around the temporal direction defined by our thermal boundary conditions). 
While our calculations are always thermally deconfined, we will be interested in exploring the `spatial confinement' phase transition between these two phases, signalled by the spatial Wilson line.

We proceed by fixing the aspect ratio $\alpha$ and simultaneously varying both $r_{\beta}$ and $r_L$.
This corresponds to scanning straight lines in the $r_{\beta}$--$r_L$ plane with slope $\alpha^{-1}$, as shown in \fig{fig:expec}.
In this way, larger aspect ratios will give access to transitions in the high-temperature region approaching the two-dimensional gauge--scalar system, while smaller aspect ratios will probe transitions in the low-temperature regime relevant to holography.
In this low-temperature holographic regime, we expect the spatial confinement transition to follow the curve described by
\begin{equation}
  \label{eq:holo}
  r_L^{(5-p)/2} = r_L^{3/2} = c_{\text{grav}} r_{\beta},
\end{equation}
where $p = 2$ is the number of spatial dimensions and $c_{\text{grav}}$ is a constant characterizing the transition~\cite{Morita:2014md, Catterall:2017lub}.
Figure~\ref{fig:expec} also includes a sketch of such a curve.
Rearranging this expression in terms of the aspect ratio, we find
\begin{equation}
  \label{eq:alpha_dep}
  \sqrt{r_{\beta} \alpha^3} = c_{\text{grav}},
\end{equation}
informing us how the interesting range of temperatures $t = 1 / r_{\beta}$ can be expected to vary with $\alpha$.

In addition to taking the continuum limit $N_L, N_t \to \infty$ with $\lalat\rightarrow 0$ (automatically removing the two deformations as discussed above), in order to investigate phase transitions we also wish to consider the thermodynamic limit.
Because we fix the dimensionless volume $V \propto r_L^2 \times r_{\beta}$ of the skewed 3-torus~\cite{Catterall:2020nmn}, the thermodynamic limit actually corresponds to the large-$N$ limit of the U($N$) gauge group.
For this reason, we will present below results for a large number of colors, $N = 8$.
This large $N$ is achieved at the cost of considering relatively small lattice volumes, at most $16^3$.

\section{\label{sec:results}Current results for aspect ratio $\alpha = 1$}
\begin{figure}[tbp]
  \centering
  \includegraphics[width=0.97\textwidth]{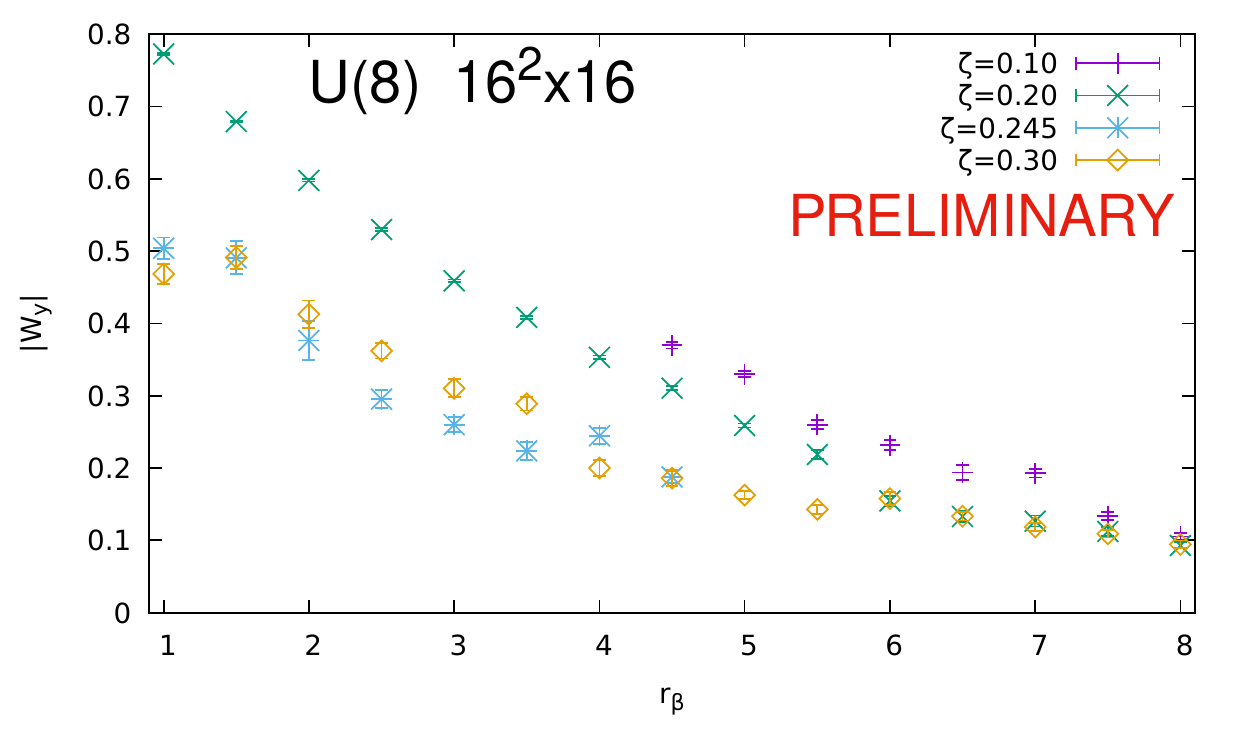}
  \caption{\label{fig:magnitude}Magnitude of the Wilson line $W$ in the $y$ direction as a function of inverse temperature $r_\beta$ for a $16^3$ lattice volume, $N = 8$ colors and several values of the deformation parameter $\zeta$. The results are consistent with the expectation $W=0$ for large $r_L = \alpha r_\beta$ and large $N$.}
\end{figure}

\begin{figure}[tbp]
  \centering
  \includegraphics[width=0.97\textwidth]{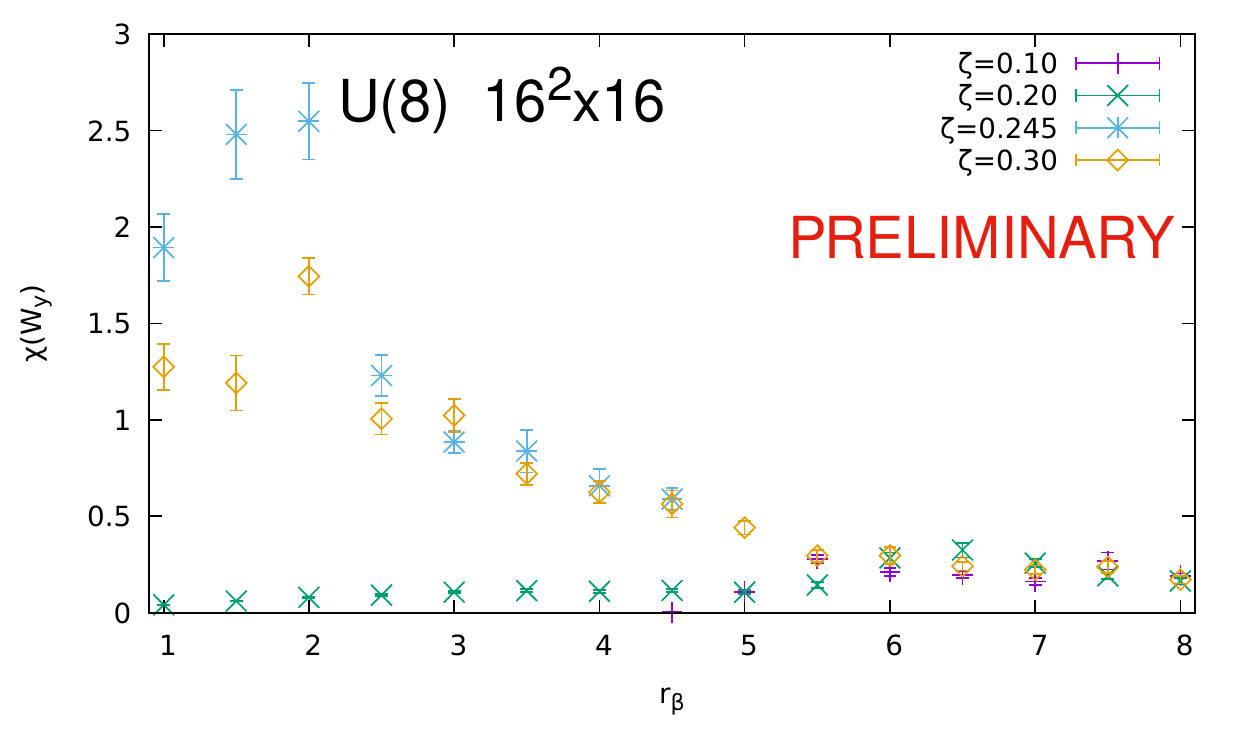}
  \caption{\label{fig:suscept}Susceptibility of the Wilson line in the $y$ direction as a function of inverse temperature $r_\beta$ for a $16^3$ lattice volume, $N=8$ colors and several values of the deformation parameter $\zeta$.}
\end{figure}

\begin{figure}[tbp]
  \centering
  \includegraphics[width=0.97\textwidth]{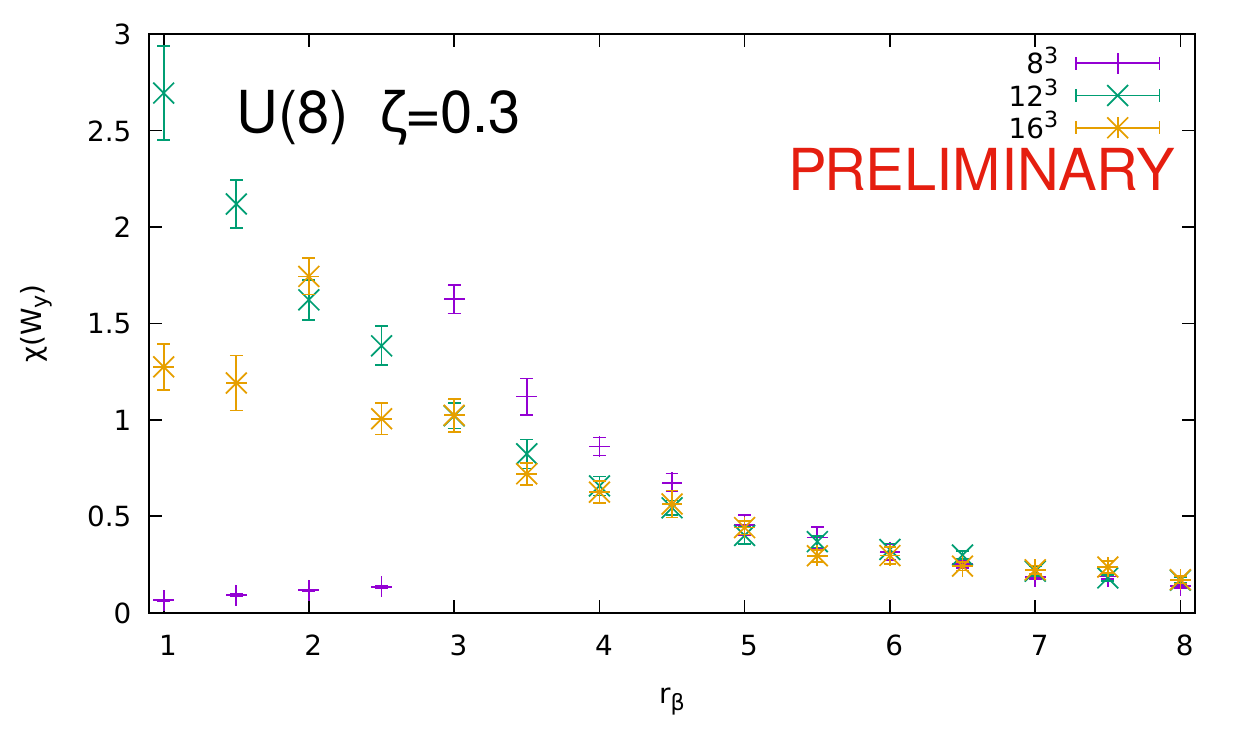}
  \caption{\label{fig:susceptzeta}Susceptibility of the Wilson line in the $y$ direction as a function of inverse temperature $r_\beta$ for deformation parameter $\zeta=0.3$, $N=8$ colors and several lattice volumes.}
\end{figure}

Fixing the aspect ratio $\alpha = 1$, corresponding to $N_L^3$ lattice volumes, Figs.~\ref{fig:magnitude}, \ref{fig:suscept} and \ref{fig:susceptzeta} present our U(8) results for the Wilson line in the $y$ direction.
We have checked that $W_z$ behaves in the same way.
Specifically, \fig{fig:magnitude} plots the magnitude $|W_y|$ for $16^3$ lattices with several different values of the parameter $\zeta$ that controls the soft-supersymmetry-breaking deformations.
Figure~\ref{fig:suscept} shows the corresponding susceptibilities, while \fig{fig:susceptzeta} fixes $\zeta = 0.3$ and compares three different lattice volumes, $N_L = 8$, $12$ and $16$.

These results are obtained from binned jackknife analyses with bin sizes set to be larger than autocorrelation times that we estimate using the `autocorr' module in \texttt{emcee}~\cite{Foreman:2013mc}.
Specifically, after thermalization/equilibration we check autocorrelation times for a fermion bilinear~\cite{Catterall:2015ira}, the temporal Polyakov loop, and the lowest eigenvalue of the fermion operator.
On $16^3$ lattices, the Polyakov loop and fermion eigenvalue typically have autocorrelation times of hundreds of MDTU.
Accounting for sometimes slow thermalization that requires thousands of MDTU, we generate up to 25,000 MDTU per ensemble, producing at least 12 bins for each jackknife analysis (and sometimes over 100 bins).

Even though our results are currently preliminary, there are several observations we can make.
First, we note that either larger lattice volumes or larger deformations (or both) are needed to stabilize RHMC configuration generation at higher temperatures (smaller $r_{\beta}$).
As reflected by Figs.~\ref{fig:magnitude} and \ref{fig:suscept}, we find that $\zeta = 0.1$ is insufficient for any $r_{\beta} \leq 4$.
Next, we see that the magnitude of the spatial Wilson line behaves as expected, increasing at small $r_L = \alpha r_{\beta}$ that corresponds to the spatially deconfined D0 phase, and (up to finite-$N$ effects) vanishing in the large-$r_L$ spatially confined D2 phase.

However, this change in the magnitude of the Wilson line is not yet accompanied by well-defined peaks in the corresponding susceptibility shown in Figs.~\ref{fig:suscept} and \ref{fig:susceptzeta}.
While there are broad peaks in some cases, other data sets evolve monotonically across the temperatures we have analyzed so far.
Recall that the peak height should be roughly the same for all $L$, scaling instead with the number of colors $N$.
Overall, it's clear that the spatial confinement transition is difficult to resolve for this aspect ratio $\alpha = 1$.
This is consistent with the two-dimensional case, for which \refcite{Catterall:2017lub} was only able to resolve the transition for $\alpha \geq 3 / 2$, with much sharper transitions for larger aspect ratios up to $\alpha = 8$.
This strongly motivates our next step of moving to larger aspect ratios, for which \eq{eq:alpha_dep} allows us to use our current $\alpha = 1$ results to choose appropriate ranges of $r_{\beta}$ to analyze.

\section{\label{sec:conc}Outlook and next steps}\label{future} 
In this proceedings we have presented ongoing work investigating the phase structure of maximally supersymmetric Yang--Mills theory in three dimensions.
With aspect ratio $\alpha=1$, our results for the Wilson line and its susceptibility as functions of temperature show signs of a phase transition.
However, it is hard to resolve the value of the critical temperature, which motivates our next steps.

First, we will fill in more data points in the region of interest, to better resolve any peaks that can be obtained with $\alpha = 1$.
Even with our current results, we can use \eq{eq:alpha_dep} to determine appropriate ranges of temperatures to scan with larger aspect ratios that are expected to produce clearer phase transition signals~\cite{Catterall:2017lub}.
Further in the future, it will be interesting to generalize our work to consider different lengths in each of the two spatial dimensions, $r_x \neq r_y$.
This would give access to different transitions, for example between the D2 phase and a phase dual to a system of D1 branes.

\vspace{20 pt} 
\noindent \textsc{Acknowledgments:}~We thank Raghav Jha, Anosh Joseph and Toby Wiseman for helpful conversations and continuing collaboration on lattice supersymmetry.
Numerical calculations were carried out at the University of Liverpool and at the University of Cambridge through the STFC DiRAC facility.
DS was supported by UK Research and Innovation Future Leader Fellowship {MR/S015418/1} and STFC grant {ST/T000988/1}.

\bibliographystyle{JHEP}
\bibliography{proc}
\end{document}